# DLSF: Dual-Layer Synergistic Fusion for High-Fidelity Image Synthesis

Zhen-Qi Chen
National Yang Ming Chiao Tung University
1001 University Road, Hsinchu 300, Taiwan
alex.ii13@nycu.edu.tw

Yuan-Fu Yang
National Yang Ming Chiao Tung University
1001 University Road, Hsinchu 300, Taiwan
yfyangd@gmail.com

**Abstract**

*With the rapid advancement of diffusion-based generative models, Stable Diffusion (SD) has emerged as a state-of-the-art framework for high-fidelity image synthesis. However, existing SD models suffer from suboptimal feature aggregation, leading to incomplete semantic alignment and loss of fine-grained details, especially in highly textured and complex scenes. To address these limitations, we propose a novel dual-latent integration framework that enhances feature interactions between the base latent and refined latent representations. Our approach employs a feature concatenation strategy followed by an adaptive fusion module, which can be instantiated as either (i) an Adaptive Global Fusion (AGF) for hierarchical feature harmonization, or (ii) a Dynamic Spatial Fusion (DSF) for spatially-aware refinement. This design enables more effective cross-latent communication, preserving both global coherence and local texture fidelity. Our GitHub page: https://anonymous.4open.science/r/MVA2025-22 .*

## 1. Introduction

Diffusion Models have emerged as one of the most influential technologies in generative deep learning, demonstrating remarkable results in various applications including image synthesis, image editing, and visual understanding [2, 9, 23, 27, 28]. Although traditional Generative Adversarial Networks (GANs) [3] and Variational Autoencoders (VAEs) [10] have achieved successes in image generation tasks, they still encounter issues such as mode collapse, training instability, and limitations in image fidelity. In contrast, Diffusion Models adopt a denoising approach that renders the image generation process more robust while excelling at preserving fine details and increasing sample diversity.

Since the introduction of Denoising Diffusion Probabilistic Models (DDPM) [4], the use of Diffusion Models has rapidly expanded to high-resolution image synthesis, text-to-image generation, medical image processing [29], and many other fields. Among these advancements, Latent Diffusion Models (LDMs) [5] perform the diffusion process within a latent space, reducing computational cost and enhancing image fidelity, thereby making high-resolution image generation more efficient. Further progress, exemplified by Stable Diffusion XL (SDXL) [6], has led to improvements in image detail representation and overall consistency, achieving a new quality benchmark in conditional control scenarios [17].

Although SDXL and other diffusion models have demonstrated outstanding capabilities in image generation, they still encounter deficiencies in latent feature fusion that lead to noticeable issues with semantic consistency and detail integrity. In particular, when dealing with highly dynamic, complex backgrounds or high-frequency textures [30, 31], SDXL remains prone to semantic misalignment or detail blurring. These problems largely stem from inadequate multi-scale information integration within the latent feature space, as most methods rely on single-level feature fusion and overlook the need for inter-level alignment, thus compromising overall image quality [25, 26]. In view of these challenges, this study proposes an efficient feature fusion strategy to enhance semantic consistency and detail retention in diffusion image generation. We contend that precise alignment and reinforcement of latent features are critical for improving image quality. Accordingly, we introduce a fusion mechanism that reinforces feature aggregation by promoting interaction between the base and refined latent representations, thereby enhancing structural integrity and detail fidelity.

To achieve this goal, we design two complementary feature fusion modules: the Adaptive Global Fusion (AGF) and the Dynamic Spatial Fusion (DSF). AGF primarily focuses on aligning and enhancing hierarchical features through cross-level information flow to improve the global consistency of generated images. Compared with the traditional Feature Pyramid Network (FPN) [7], AGF places greater emphasis on adaptive cross-level feature adjustments to effectively reduce semantic misalignment while maintaining detail fidelity. Building on this foundation, DSF adopts the spatial attention mechanism of the Convolutional Block Attention Module (CBAM) [8] and optimizes it for the feature distribution in diffusion models, ensuring robust preservation and reinforcement of high-frequency information throughout the diffusion process. By integrating these two modules, our approach not only addresses the limitations of existing models in semantic alignment but also strengthens local details at every stage of image generation, ultimately producing images that appear more natural and coherent [32, 24].

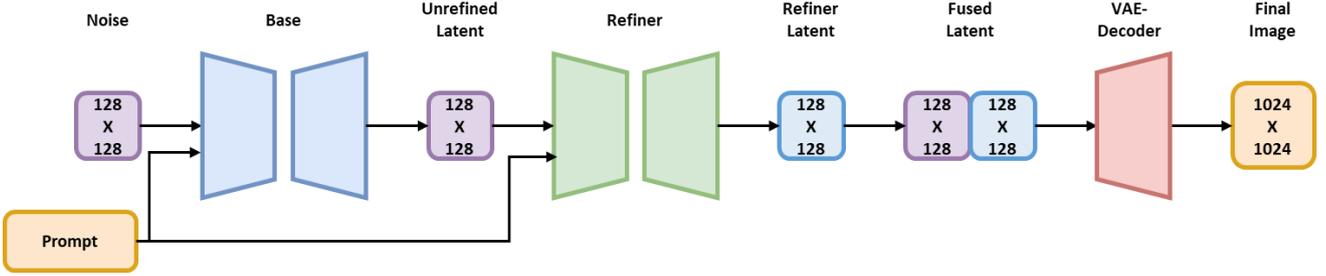

Figure 1. The pipeline of DLSF. The pipeline consists of five stages for high-quality multi-view image synthesis. First, the prompt encoding embeds the textual input into a latent space. The base model generates the base latent representation (128×128), capturing global structural features, while the refine model enhances it to produce the refined latent representation with finer details. These representations are fused using DLSF, employing either Adaptive Global Fusion or Dynamic Spatial Fusion to produce the fused latent representation. Finally, the decoder generates the high-resolution (1024×1024) multi-view images, balancing structural consistency and detailed reconstruction.

## 2. Method

### 2.1 Preliminaries

SDXL represents a significant advancement in LDMs, a family of generative models that execute the diffusion process in a compressed latent space. This approach synergizes denoising score matching [4] with VAEs techniques, enabling the synthesis of high-resolution images with significantly reduced computational costs compared to pixel-space diffusion.

Earlier models like Stable Diffusion [5] showcased the potential of LDMs high-quality image generation, incorporating innovative U-Net architectures [12] with skip connections and attention layers [1], as well as mechanisms for conditioning on text or other modalities [11, 15]. Stable Diffusion also emphasized the importance of scalable training on large-scale datasets like LAION-400M [14], facilitating robust and diverse model capabilities. Building upon these foundational concepts, SDXL integrates a range of architectural and training advancements to push the boundaries of LDMs performance. According to Podell et al. [6], one of SDXL's key innovations is the adoption of a threefold larger U-Net architecture, which leverages deeper layers and a significantly expanded capacity for feature extraction and generation. To improve conditional control, SDXL introduces a dual text encoder system, which simultaneously processes prompts at multiple semantic levels, addressing limitations in fine-grained alignment between input text and output images.

Furthermore, SDXL implements micro-conditioning techniques, a refinement of traditional conditioning that allows for finer control over latent variables, ensuring spatial and semantic coherence even in complex scenes. Podell et al. [6] also highlighted the importance of multi-scale training strategies, which involve training models across varying resolutions to address challenges related to inconsistent image resolutions during inference. These strategies enhance the model's adaptability to diverse input-output conditions, further improving robustness.

Finally, SDXL incorporates a diffusion-based post-processing module [16], which applies additional refinement to synthesized images, enhancing texture consistency and structural accuracy. This step is particularly effective in use cases demanding high fidelity, such as medical imaging [29], fine-grained art generation [2], and other high-precision applications.

### 2.2 DLSF

The proposed system employs a two-stage approach for multi-view image synthesis. First, the base model processes a single input image to generate a base latent ($L_b$), capturing the global features of the image. Next, the refine model further processes the base latent to extract detailed features, producing a refined latent ($L_r$), The base latent and refined latent are then combined to form a fused latent representation ($L_f$), which is fed into the Decoder to generate multi-view images. By combining global and detailed features, the system achieves high-quality multi-view outputs, enhancing structural consistency and detail reconstruction capabilities. Figure 1 illustrates our pipeline.

#### 2.2.1 Adaptive Global Fusion

To effectively combine the global and detailed features captured by the base latent ($L_b$) and refined latent ($L_r$), we introduce an Adaptive Global Fusion. This module dynamically allocates attention weights to each latent representation to emphasize the most relevant information for multi-view image synthesis.

Given $L_b \in \mathbb{R}^{C \times H \times W}$ where $C$, $H$, and $W$ represent the number of channels, height, and width respectively, the two latent representations are concatenated along the channel dimension to form a combined input:

$$L_{concat} = Concat(L_b, L_r), \qquad L_{concat} \in \mathbb{R}^{2C \times H \times W}$$

This concatenated tensor is processed by a 1×1 convolutional layer to generate attention logits:

$$A = Conv_{7 \times 7}(L_{concat}), \qquad A \in \mathbb{R}^{2 \times H \times W}$$

The attention logits are normalized using a softmax operation along the channel dimension to produce attention weights:

$$W = Softmax(A), \quad W \in \mathbb{R}^{2 \times H \times W}$$

$$W_b(h,w) = W(0,h,w), \qquad W_r(h,w) = W(1,h,w)$$

The final fused latent representation is computed as the weighted sum of the base latent and refined latent:

$$L_f = W_b \odot L_b + W_r \odot L_r$$

This fused latent representation effectively balances global and detailed features, ensuring structural consistency and high-fidelity detail reconstruction.

### 2.2.2 Dynamic Spatial Fusion

To further enhance the fusion of latent representations, we propose a Dynamic Spatial Fusion, which leverages spatial attention maps to dynamically allocate importance across the spatial dimensions of the latent representations.

For each pair of $L_b$ and $L_r$, we first extract spatial features using average pooling and max pooling along the channel dimension:

$$P_{avg} = AvgPool(L_r), \qquad P_{max} = MaxPool(L_b),$$

$$P_{avg}, P_{max} \in \mathbb{R}^{1 \times H \times W}$$

These pooled features are concatenated to form a spatial feature tensor:

$$P_{concat} = Concat(P_{avg}, P_{max})$$

The spatial feature tensor is processed by a 7×7 convolutional layer to generate the spatial attention map:

$$M_{spatial} = Sigmoid\left(Conv_{7 \times 7}(P_{spatial})\right),$$

$$M_{spatial} \in \mathbb{R}^{1 \times H \times W}$$

The final fused latent representation is computed as a spatially weighted combination of the base latent and refined latent:

$$L_f = M_{spatial} \odot L_r + (1 - M_{spatial}) \odot L_b$$

This approach dynamically adjusts the contribution of each latent representation at a pixel level, enabling finer control over feature integration.

## 3. Experiments

### 3.1 Implementation Details

We utilize the SDXL architecture, which includes a base model (stable-diffusion-xl-base-1.0) and a refiner model (stable-diffusion-xl-refiner-1.0). Both models process latent tensors of size (1, 4, 128, 128) to generate high-quality images. The final output images are at a resolution of 1024×1024 pixels.

The models are implemented using PyTorch [21] and the Diffusers library, with half-precision (FP16) for optimized memory usage and faster computation. Inference is performed on an NVIDIA A6000 GPU with 48 GB VRAM, ensuring efficient handling of the SDXL pipeline.

### 3.2 Results and Comparisons

To thoroughly evaluate the effectiveness of our proposed AGF and DSF for large-scale conditional image generation, we conducted extensive experiments on the ImageNet [18] dataset at resolutions of 256×256 and 512×512. These two modules address different but complementary aspects: AGF focuses on synergizing latent representations to enhance overall structural consistency, whereas DSF employs refined spatial attention mechanisms to preserve fine-grained details and textures.

We adopt SDXL as our baseline. The base model is first run for 50 DDIM steps [19] to obtain initial latent representations, followed by an additional 15 DDIM steps using the refine model for further refinement. We set the Classifier-Free Guidance (CFG) parameter to 5 [13], balancing the trade-off between diversity and realism. In total, we generate 5,000 images spanning 1,000 classes (5 images per class) to fully assess the generalization capability and robustness of our approach. Quantitative results are shown in Table 1 and Table 2, comparing our method with the SDXL baseline using standard metrics, while qualitative results are presented in Figure 2.

Experimental results demonstrate that both AGF and DSF outperform the SDXL baseline at 256×256 and 512×512 resolutions, achieving notably lower Fréchet Inception Distance FID [20] and sFID scores. For instance, at 256×256, AGF and DSF achieve FID scores of 18.79 and 18.89, respectively, surpassing SDXL's 20.16; at 512×512, both modules reach an FID of 18.70, outperforming SDXL's 19.65. Moreover, the Inception Score IS [22] indicates significant improvements in perceptual quality and diversity: at 256×256, DSF attains 232.04, exceeding AGF 230.43 and SDXL 219.74; at 512×512, AGF and DSF achieve 243.48 and 243.62, respectively, substantially outperforming SDXL 234.75. In terms of Precision and Recall, AGF and DSF further enhance the balance between realism and coverage, raising Precision from 0.86 to 0.87 and Recall from 0.35 to 0.39 at the 256×256 resolution—indicating the capacity to produce a broader and higher-quality sample distribution.

Overall, our results confirm that AGF and DSF effectively strengthen SDXL-based generative modeling by simultaneously improving global structural coherence and local detail preservation, leading to marked gains in multiple evaluation metrics. These findings underscore the potential of our approach to advance the state of conditional image generation, offering substantial benefits in visual quality, diversity, and structural integrity, and providing a comprehensive solution for future research.

Table 1. Comparison of our methods with the SDXL baseline model on Class-Conditional ImageNet at 256×256 resolution.

| Class-Conditional ImageNet 256x256 | | | | | |
|---|---|---|---|---|---|
| Method | FID↓ | sFID↓ | IS↑ | Prec↑ | Rec↑ |
| SDXL | 20.16 | 48.98 | 219.74 | 0.86 | 0.35 |
| AGF | 18.79 | 47.64 | 230.43 | 0.87 | 0.39 |
| DSF | 18.89 | 48.21 | 232.04 | 0.87 | 0.39 |

Table 2. Comparison of our methods with the SDXL baseline model on Class-Conditional ImageNet at 512×512 resolution.

| Class-Conditional ImageNet 512x512 | | | | | |
|---|---|---|---|---|---|
| Method | FID↓ | sFID↓ | IS↑ | Prec↑ | Rec↑ |
| SDXL | 19.65 | 50.54 | 234.75 | 0.86 | 0.35 |
| AGF | 18.70 | 49.77 | 243.48 | 0.85 | 0.38 |
| DSF | 18.70 | 50.22 | 243.62 | 0.85 | 0.38 |

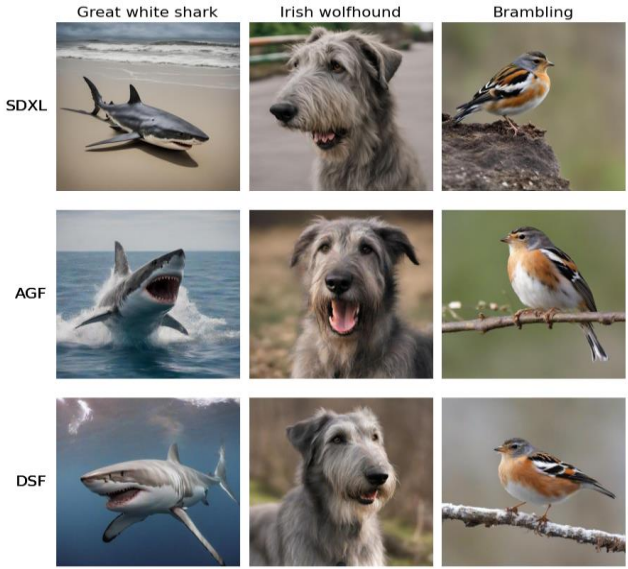

Figure 2. Qualitative comparison among SDXL, AGF, and DSF on three representative ImageNet classes. SDXL's outputs exhibit notable flaws, such as placing the shark on a sand beach, distorting the wolfhound's mouth, and omitting a leg on the bird. In contrast, AGF and DSF preserve more accurate structures and coherent details.

### 3.3 Ablation studies

To investigate the impact of different latent feature fusion strategies on image generation quality, we conduct an ablation study on the Class-Conditional ImageNet 512×512 dataset. This study compares our proposed AGF and DSF, and further analyzes whether applying an additional refine model step after fusing the base latent and refined latent (denoted as /r) provides further improvements.

As shown in Table 3, both AGF/r and DSF/r exhibit an increase in FID and sFID, indicating that the additional refinement step may lead to excessive smoothing or unnecessary modifications to local details. Moreover, the IS decreases, suggesting that this step may reduce image diversity and limit the variation in generated content. Specifically, AGF/r shows a noticeable drop in IS from 243.48 to 215.06, and DSF/r exhibits a decline from 243.62 to 218.36, further validating that additional refinement may affect image diversity.

Furthermore, although Precision and Recall values do not vary significantly across methods, both AGF and DSF outperform their refined versions, indicating that latent feature fusion indeed improves image generation quality, while additional refinement does not necessarily confer further benefits and may instead cause detail loss and reduced diversity. These findings underscore the critical balance between effective fusion and the risk of over-smoothing.

Table 3. Ablation study results on latent feature fusion strategies. We compares different latent feature fusion strategies and analyzes the impact of applying an additional refine model (/r) before decoding on image generation quality.

| Class-Conditional ImageNet 512x512 | | | | | |
|---|---|---|---|---|---|
| Method | FID↓ | sFID↓ | IS↑ | Prec↑ | Rec↑ |
| AGF | 18.70 | 49.77 | 243.48 | 0.852 | 0.381 |
| AGF /r | 20.02 | 53.94 | 215.06 | 0.851 | 0.363 |
| DSF | 18.70 | 50.22 | 243.62 | 0.854 | 0.383 |
| DSF /r | 19.89 | 54.28 | 218.36 | 0.853 | 0.383 |

## 4. Conclusion

This paper presents the DLSF approach, a novel high-fidelity image synthesis and feature fusion framework designed to significantly enhance the generative performance of diffusion models, particularly SDXL. By integrating the AGF and DSF, DLSF effectively fuses Base Latent and Refined latent representations, achieving multi-level feature integration while preserving both global structure and local texture details. Experimental results demonstrate that DLSF outperforms SDXL in terms of both realism and diversity, clearly underscoring its potential to elevate overall generative quality. Furthermore, our ablation studies reveal that excessive refinement following latent feature fusion can lead to undesirable smoothing effects and diminished diversity, highlighting the critical importance of maintaining a precisely controlled fusion mechanism. Looking ahead, we plan to explore more advanced attention strategies—such as adaptive weight allocation and dynamic feature selection—to further optimize latent fusion. We also aim to extend DLSF's applicability to specialized domains like medical imaging and scientific visualization. Overall, our approach not only advances general image synthesis but also promises to serve as a cornerstone for next-generation generative models, driving applications in complex scenarios and fostering cross-disciplinary innovation.

# References

[1] Vaswani, A. et al.: "Attention Is All You Need," Advances in Neural Information Processing Systems, 2017. 2

[2] Zhang, C., Zhang, C., Zhang, M., and I. S. Kweon: "Text-to-Image Diffusion Models in Generative AI: A Survey," arXiv preprint arXiv:2303.07909, 2023. 1, 2

[3] Goodfellow, I., Pouget-Abadie, J., Mirza, M., Xu, B., Warde-Farley, D., Ozair, S., et al.: "Generative Adversarial Nets," Advances in Neural Information Processing Systems, vol. 27, 2014. 1

[4] Ho, J., Jain, A., and Abbeel, P.: "Denoising Diffusion Probabilistic Models," Advances in Neural Information Processing Systems, vol. 33, pp. 6840–6851, 2020. 1, 2

[5] Rombach, R., Blattmann, A., Lorenz, D., Esser, P., and Ommer, B.: "High-Resolution Image Synthesis with Latent Diffusion Models," in Proc. IEEE/CVF Conference on Computer Vision and Pattern Recognition, pp. 10684–10695, 2022. 1, 2

[6] Podell, D., English, Z., Lacey, K., Blattmann, A., Dockhorn, T., Müller, J., et al.: "SDXL: Improving Latent Diffusion Models for High-Resolution Image Synthesis," arXiv preprint arXiv:2307.01952, 2023. 1, 2

[7] Lin, T. Y., Dollár, P., Girshick, R., He, K., Hariharan, B., and Belongie, S.: "Feature Pyramid Networks for Object Detection," in Proc. IEEE Conference on Computer Vision and Pattern Recognition, pp. 2117–2125, 2017. 1

[8] Woo, S., Park, J., Lee, J. Y., and Kweon, I. S.: "CBAM: Convolutional Block Attention Module," in Proc. European Conference on Computer Vision (ECCV), pp. 3–19, 2018. 1

[9] Sohl-Dickstein, J., Weiss, E., Maheswaranathan, N., and Ganguli, S.: "Deep Unsupervised Learning Using Nonequilibrium Thermodynamics," in Proc. International Conference on Machine Learning, pp. 2256–2265, 2015, PMLR. 1

[10] Kingma, D. P.: "Auto-Encoding Variational Bayes," arXiv preprint arXiv:1312.6114, 2013. 1

[11] Nichol, A., Dhariwal, P., Ramesh, A., Shyam, P., Mishkin, P., McGrew, B., et al.: "GLIDE: Towards Photorealistic Image Generation and Editing with Text-Guided Diffusion Models," arXiv preprint arXiv:2112.10741, 2021. 2

[12] Ronneberger, O., Fischer, P., and Brox, T.: "U-Net: Convolutional Networks for Biomedical Image Segmentation," in Proc. MICCAI 2015, Part III, pp. 234–241, Springer International Publishing, 2015. 2

[13] Ho, J. and Salimans, T.: "Classifier-Free Diffusion Guidance," arXiv preprint arXiv:2207.12598, 2022. 3

[14] Schuhmann, C., Vencu, R., Beaumont, R., Kaczmarczyk, R., Mullis, C., Katta, A., et al.: "LAION-400M: Open Dataset of CLIP-Filtered 400 Million Image-Text Pairs," arXiv preprint arXiv:2111.02114, 2021. 2

[15] Saharia, C., Chan, W., Saxena, S., Li, L., Whang, J., Denton, E. L., et al.: "Photorealistic Text-to-Image Diffusion Models with Deep Language Understanding," Advances in Neural Information Processing Systems, vol. 35, pp. 36479–36494, 2022. 2

[16] Meng, C., He, Y., Song, Y., Song, J., Wu, J., Zhu, J. Y., and Ermon, S.: "SDedit: Guided Image Synthesis and Editing with Stochastic Differential Equations," arXiv preprint arXiv:2108.01073, 2021. 2

[17] Esser, P., Rombach, R., and Ommer, B.: "Taming Transformers for High-Resolution Image Synthesis," in Proc. IEEE/CVF Conference on Computer Vision and Pattern Recognition, pp. 12873–12883, 2021. 1

[18] Deng, J., Dong, W., Socher, R., Li, L. J., Li, K., and Fei-Fei, L.: "ImageNet: A Large-Scale Hierarchical Image Database," in Proc. IEEE Conference on Computer Vision and Pattern Recognition, pp. 248–255, 2009. 3

[19] Song, J., Meng, C., and Ermon, S.: "Denoising Diffusion Implicit Models," arXiv preprint arXiv:2010.02502, 2020. 3

[20] Heusel, M., Ramsauer, H., Unterthiner, T., Nessler, B., and Hochreiter, S.: "GANs Trained by a Two Time-Scale Update Rule Converge to a Local Nash Equilibrium," Advances in Neural Information Processing Systems, vol. 30, 2017. 3

[21] Paszke, A., Gross, S., Massa, F., Lerer, A., Bradbury, J., Chanan, G., et al.: "PyTorch: An Imperative Style, High-Performance Deep Learning Library," Advances in Neural Information Processing Systems, vol. 32, 2019. 3

[22] Salimans, T., Goodfellow, I., Zaremba, W., Cheung, V., Radford, A., and Chen, X.: "Improved Techniques for Training GANs," Advances in Neural Information Processing Systems, vol. 29, 2016. 3

[23] Hu, X., Jin, Y., Liang, J., Liu, J., Luo, R., Li, M., and Peng, T.: "Diffusion Model for Image Generation—A Survey," in Proc. 2023 2nd International Conference on Artificial Intelligence, Human-Computer Interaction and Robotics (AIHCIR), pp. 416–424, IEEE, Dec. 2023. 1

[24] Ho, J., Saharia, C., Chan, W., Fleet, D. J., Norouzi, M., and Salimans, T.: "Cascaded Diffusion Models for High Fidelity Image Generation," Journal of Machine Learning Research, vol. 23, no. 47, pp. 1–33, 2022. 1

[25] Ye, B., Xue, R., and Wu, Q.: "A Hybrid Attention Multi-Scale Fusion Network for Real-Time Semantic Segmentation," Scientific Reports, vol. 15, no. 1, p. 872, 2025. 1

[26] Cui, H., Li, J., Hua, Z., and Fan, L.: "Attention-Guided Multi-Scale Feature Fusion Network for Low-Light Image Enhancement," Frontiers in Neurorobotics, vol. 16, p. 837208, 2022. 1

[27] Huang, Y., Huang, J., Liu, Y., Yan, M., Lv, J., Liu, J., et al.: "Diffusion Model-Based Image Editing: A Survey," arXiv preprint arXiv:2402.17525, 2024. 1

[28] Yang, L., Zhang, Z., Song, Y., Hong, S., Xu, R., Zhao, Y., et al.: "Diffusion Models: A Comprehensive Survey of Methods and Applications," ACM Computing Surveys, vol. 56, no. 4, pp. 1–39, 2023. 1

[29] Shi, Y., Abulizi, A., Wang, H., Feng, K., Abudukelimu, N., Su, Y., and Abudukelimu, H.: "Diffusion Models for Medical Image Computing: A Survey," Tsinghua Science and Technology, vol. 30, no. 1, pp. 357–383, 2024. 2

[30] Gatys, L., Ecker, A. S., and Bethge, M.: "Texture Synthesis Using Convolutional Neural Networks," Advances in Neural Information Processing Systems, vol. 28, 2015. 1

[31] Johnson, J., Alahi, A., and Fei-Fei, L.: "Perceptual Losses for Real-Time Style Transfer and Super-Resolution," in Proc. Computer Vision–ECCV 2016, Part II, pp. 694–711, Springer International Publishing, 2016. 1

[32] Dai, Y., Gieseke, F., Oehmcke, S., Wu, Y., and Barnard, K.: "Attentional Feature Fusion," in Proc. IEEE/CVF Winter Conference on Applications of Computer Vision, pp. 3560–3569, 2021. 1